**Intermediate range order in (Fe,Al) silicate network glasses: a neutron diffraction and EPSR modeling investigation.**

C. Weigel,[1] L. Cormier[*,1] G. Calas,[1] L. Galoisy,[1] and D.T. Bowron[2]

[1]*Institut de Minéralogie et de Physique des Milieux Condensés, Université Pierre et Marie Curie-Paris 6, Université Denis Diderot, CNRS UMR 7590, IPGP, 4 place Jussieu, 75005 Paris, France*

[2]*ISIS Facility, CCLRC Rutherford Appleton Laboratory, Chilton, Didcot, Oxon OX11 OQX, UK*

[*]Corresponding author: cormier@impmc.jussieu.fr

## ABSTRACT

The local structural environment and the spatial distribution of iron and aluminum ions in sodosilicate glasses with composition $NaFe_xAl_{1-x}Si_2O_6$ (x = 1, 0.8, 0.5 and 0) is studied by high-resolution neutron diffraction combined with structural modeling using the Empirical Potential Structure Refinement (EPSR) code. This work gives evidence of differences in the structural behavior of $Al^{3+}$ and $Fe^{3+}$, which are both often considered to act as network formers in charge-balanced compositions. The short-range environment and the structural role of the two cations are not composition dependent, and hence the structure of intermediate glasses can then be seen as a mixture of the structures of the two end-members. All $Al^{3+}$ is 4-coordinated for a distance $d_{{}^{[4]}Al^{3+}-O} = 1.76 \pm 0.01 Å$. The high-resolution neutron data allows deciphering between two populations of Fe. The majority of $Fe^{3+}$ is 4-coordinated ($d_{{}^{[4]}Fe^{3+}-O} = 1.87 \pm 0.01 Å$) while the remaining $Fe^{3+}$ and all $Fe^{2+}$ (~12% of total Fe) are 5-coordinated ($d_{{}^{[5]}Fe-O} = 2.01 \pm 0.01 Å$). Both $AlO_4$ and $FeO_4$ are randomly distributed and connected with the silicate network in which they share corners with $SiO_4$ tetrahedra, in agreement with a network-forming role of those species. On the contrary $FeO_5$ tends to form clusters and to share edges with each other. 5-coordinated Fe is interpreted as network modifier and it turns out that, even if this coordination number is rare in crystals, it is more common in glasses in which they can have a key role on physical properties.

# I. INTRODUCTION

Network glasses, such as silica or silicate glasses, are amorphous materials of major technological and geophysical importance. These glasses consist of a tetrahedral framework of corner-sharing tetrahedra building rings of various sizes. Such as in other amorphous and liquid systems, medium range order (MRO) continues to attract attention.[1, 2] Indeed the organization at this scale determines physical properties such as chemical diffusion, electrical conductivity, magnetic properties… The determination of MRO in glasses and melts is difficult due to the superimposition of the different atomic pairs beyond 3 Å. Numerical modeling of glass structure based on experimental data, such as Reverse Monte Carlo (RMC) or Empirical Potential Structure Refinement (EPSR) methods, are valuable tools to obtain structural information on medium- and extended-range structure of glasses through the intimate connection between computation and experiment.

Among network glasses, aluminosilicate glasses have been extensively studied as structural analogues of amorphous silica. In aluminosilicate glasses, Al is 4-coordinated and no detectable amount of high coordinated Al has been detected in alkali aluminosilicate glasses at room pressure.[3-6] For Al/Na=1, these glasses retain a pure 3D framework topology without significant clustering of Al.[7] Multicomponent glasses are interesting in showing the influence of the individual glass components on the 3D framework topology through a substitution of elements with a similar structural behavior. Recent investigations have shown the interest of Fe-based framework glasses, in scaling physical properties of silica and 3D-silicate glasses.[8] However, as $Fe^{3+}$ decreases the viscosity of silicate melts, relative to their aluminosilicate analogues,[9] there is evidence of the presence of non-tetrahedral $Fe^{3+}$ species that may be retained in the Fe-bearing glasses quenched from these melts. The presence of 5-coordinated $Fe^{3+}$ ($^{[5]}Fe^{3+}$) has been recently shown in a $NaFeSi_2O_6$ glass .[10] Owing to the differences between Al- and Fe-bearing glasses, it is interesting to investigate the influence of the Fe/Al mixing on the structure of a framework glass and the clustering of these cations. Indeed, a deviation from a pure 3D-glass structure compositions more dilute in Fe content and the associated clustering of cations may explain peculiar spin-lattice relaxation times or optical properties of Fe-bearing silicate glasses, which suggest that $Fe^{3+}$ ions are distributed in close enough proximity to another $Fe^{3+}$ even at modest Fe concentrations.[11, 12]

We present experimental and numerical data on the influence of the Fe for Al substitution on the structure of amorphous $NaAlSi_2O_6$, a model framework glass. The combination of neutron diffraction and numerical modeling show the influence of Al and Fe on the intermediate range organization of these glasses. Whereas Al is randomly distributed and shares only corners with the other cationic polyhedra, there is direct evidence of Fe clustering. This Fe clustering is observed even at low Fe substitution levels (2 at% Fe). It explains the peculiar optical properties by intervalence charge transfer that can take place between adjacent $Fe^{2+}$ and $Fe^{3+}$.[13] It can also explain electronic conduction process by electron hopping between the neighboring $Fe^{2+}$ and $Fe^{3+}$ in the clusters.[14, 15] Assuming that the structure of the glass is retained in the liquid state, our structural model can also explain the low viscosity of Fe-bearing molten silicate frameworks.[16, 17]

## II. EXPERIMENTAL PROCEDURES

### A. Materials preparation

Four glasses, of composition $NaFe_xAl_{1-x}Si_2O_6$ (x = 1, 0.8, 0.5 and 0), were prepared from stoichiometric mixtures of dried reagent grade $SiO_2$, $Al_2O_3$, $Fe_2O_3$ and $Na_2CO_3$. The following denomination of samples will be used: NFS for x = 1, NFA0.8 for x = 0.8, NFA0.5 for x = 0.5 and NAS for x = 0. The starting powder mixtures were decarbonated at 750°C during 12 h in platinum crucibles. The starting materials were melted at 1100°C in an electric furnace in air for 2h. The temperature was then brought to 1300°C for 2h and finally to 1450°C (1600°C for NAS) for 30 min. The melts were then quenched by rapid immersion of the crucible in water, ground to a powder and re-melted. To ensure a good chemical homogeneity, this grinding-melting process was repeated three times.

Fe-bearing samples were dark brown and appeared bubble-free. NAS was colorless and it was not possible to get rid of the presence of some bubbles, leading to a slight underestimate of its density. Qualitatively, the viscosity of melts decreases as Fe-content increases. The samples are optically isotropic under polarized light. Transmission electron microscope images confirmed the absence of nm-size heterogeneities (crystalline or amorphous). Glass composition was determined using a Cameca SX50 electron microprobe at the CAMPARIS analytical facility (Paris 6 University, France) (Table 1). Densities (Table 1) were measured by Archimedes method, with toluene as liquid reference and the redox state, defined as the relative abundance of $Fe^{3+}$, $Fe^{3+}/Fe_{tot}$, (Table 1) was determined by Mössbauer spectroscopy (BGI, Germany). All glasses along the join are oxidized, with $Fe^{3+}/Fe_{tot}$ = 88 ± 1 %, depending on the glass composition, contrary to the study of Mysen and Virgo[18] where the samples containing less than 10 at% Fe were completely reduced at 1 atm and 1450°C. As Fe may exist as $Fe^{3+}$ and $Fe^{2+}$, it is important to point out that high redox values were achieved in the compositions investigated, minimizing the influence of $Fe^{2+}$.

## B. Neutron elastic scattering experiments

Neutron elastic scattering experiments were performed at room temperature at the ISIS (Rutherford Appleton Laboratory, UK) spallation neutron source on the SANDALS diffractometer. The time-of-flight diffraction mode gives access to a wide Q-range: 0.3-50$Å^{-1}$. The samples were crushed and poured in a flat TiZr cell. To obtain a good signal to noise ratio, measurements of the samples were performed during 12h. Additional shorter measurements were carried out on the vacuum chamber, on the empty can, and on a vanadium reference. Instrument background and scattering from the sample container were subtracted from the data. Data were merged, reduced, corrected for attenuation, multiple scattering and Placzeck inelasticity effects using the Gudrun program, which is based on the codes and methods of the widely used ATLAS package.[19]

The quantity measured in a neutron diffraction experiment is the total structure factor F(Q). It can be written in the Faber-Ziman formalism[20-22] as follows:

$$F(Q) = \sum_{\alpha,\beta=1}^{n,n} c_\alpha c_\beta \overline{b_\alpha} \overline{b_\beta} \left[ A_{\alpha\beta}(Q) - 1 \right] \quad (1)$$

where $n$ is the number of distinct chemical species, $A_{\alpha\beta}(r)$ are the Faber-Ziman partial structure factors, $c_\alpha$ and $c_\beta$ are the atomic concentrations of element $\alpha$ and $\beta$, and $b_\alpha$ and $b_\beta$ are the coherent neutron scattering lengths.

The differential correlation function, D(r), is obtained from the Fourier transform of the total structure factor F(Q). D(r) is linked to the individual distribution functions $g_{\alpha\beta}(r)$ by the weighted sum:

$$D(r) = 4\pi r \rho_0 \sum_{\alpha,\beta=1}^{n} c_\alpha c_\beta \bar{b}_\alpha \bar{b}_\beta \left[ g_{\alpha\beta}(r) - 1 \right] \quad (2)$$

The neutron weighting factors for each atomic pair in the total structure factors are given in Table 2. They allow comparing the contributions of the different atomic pairs in the scattering data.

## C. Structural modeling

Since all the partial pair distribution functions are superimposed beyond 2Å, the glass structure was simulated using the Empirical Potential Structure Refinement (EPSR) code in order to extract more detailed structural information about the Fe environment and the topology of the silicate network. EPSR presents the advantage to model atomic interactions using a potential (Coulomb-Lennard Jones type) instead of the semi-hard sphere model used in RMC.

This method allows developing a structural model for liquids or amorphous solids for which diffraction data are available. It consists in refining a starting interatomic potential and atomic positions to produce the best possible agreement between the simulated and the measured structure factors.[23] The modeling was run with cubic boxes containing 4000 ions (Table 3). The dimension of the boxes (Table 3) was calculated so that the density within the boxes corresponds to the measured density. The starting potential between atom pairs was a combination of Lennard-Jones and Coulomb potentials. The potential between atoms a and b can be represented by:

$$U_{ab}(r) = 4\varepsilon_{ab}\left[\left(\frac{\sigma_{ab}}{r}\right)^{12} - \left(\frac{\sigma_{ab}}{r}\right)^{6}\right] + \frac{1}{4\pi\varepsilon_0}\frac{q_a q_b}{r} \qquad (3)$$

where $\varepsilon_{ab} = \sqrt{\varepsilon_a \varepsilon_b}$ , $\sigma_{ab} = 0.5(\sigma_a + \sigma_b)$ and $\varepsilon_0$ is the permittivity of empty space. The Lennard-Jones ε and σ values were adjusted repeatedly for NFS and NAS until the first peak in Si-O, Fe-O, Al-O and Na-O radial distribution functions occurs at about 1.63Å, 1.89Å, 1.76Å and 2.30Å respectively. Those interatomic distances were determined by Gaussian fit of the first peak of the differential correlation functions of NFS and NAS, except the value for $d_{Na\text{-}O}$, which corresponds to the interatomic distance reported in literature.[24-26] The reduced depths (ε) and effective charges[23] were used for the reference potential of those simulations and are listed in Table 4. The simulations were run at 1000K and were performed in three steps to obtain the final atomic configurations. The first step consists in refining atomic positions using only the reference potential until it reaches equilibrium, i.e. until the energy of the simulation goes to some constant value. Then, empirical potential refinement procedure is started: the empirical potentials are refined at the same time as atomic positions, in order to decrease the difference between simulated and experimental F(Q). The last step is to get averaged information. Four models were run for each composition to ensure reproducibility and increase the statistics. The results presented below are averages of those different models.

## III. RESULTS

### A. Structure factors

The total structure factors, F(Q) (Fig. 1), exhibit excellent signal-to-noise ratio up to 35 $Å^{-1}$, giving a good resolution in the real space. The structural oscillations extend up to 35 $Å^{-1}$, indicating a well-defined short-range order along the join. The main effects of the substitution of Fe for Al are observed below 11 $Å^{-1}$ and particularly for the three first peaks, showing that this substitution affects the medium range organization of the silicate network. The first peak is shifted to low Q values as Fe content decreases. The intensity of the second and the third peaks increases as Fe content decreases. The presence of isobestic points (insert in Fig. 1), regularly spaced (1.3 ± 0.2 Å), shows that the F(Q) functions of the intermediate glasses are the weighted sum between those of the two end-members, NFS and NAS. The structure of the intermediate glasses NFA0.8 and NFA0.5 is then a mixture between the structures of the end-member glasses NFS and NAS. $F_{NAS}(Q)$ is similar, especially at low Q values, to the structure factor of $NaAlSi_3O_8$ glass.[27] The medium range organization of these two sodium aluminosilicate glasses, which are typical examples of framework glasses, should then be similar.

The so-called First Sharp Diffraction Peak (FSDP) is shifted towards lower Q values as Fe is substituted for Al. Its position was determined by a Gaussian fit adjusted on its low Q side and lying on a horizontal background. FSDP is located at $Q_P$=1.75 ± 0.02 Å in $F_{NFS}(Q)$ and this position is linearly shifted to low Q values as Fe content decreases down to $Q_P$=1.62 ± 0.02 Å in $F_{NAS}(Q)$ (Fig. 2). Even if its origin is strongly debated,[28, 29] FSDP can be undoubtedly assigned to the MRO of the glass: $Q_P$ is associated with density fluctuations over a repeat distance $D = 2\pi / Q_P$, with an uncertainty on D given by $\sigma(D) = \frac{2\pi\sigma(Q_P)}{Q_P^2}$,[25] where $\sigma(Q_P)$ is the uncertainty on the position of the first peak. The characteristic repeat distance D increases from D = 3.59 ± 0.04 Å to 3.89 ± 0.05 Å in NFS and NAS glass, respectively. $Fe^{3+}$ brings then a structural ordering at lower distance than Al.

### B. Pair correlation functions

The differential correlation functions (Fig. 3) show significant differences among the glasses investigated. Indeed, the neutron scattering length of Fe being higher than the one of Al, the atomic pairs involving Fe give more intense contributions than those involving Al. The first maximum, assigned to Si-O contributions, has a Gaussian shape, with $d_{Si\text{-}O}$ = 1.63 ± 0.01 Å and $CN_{Si\text{-}O}$ = 3.9 ± 0.1, consistent with the presence of $SiO_4$ tetrahedra. Besides, this first maximum is not shifted as Fe/Al ratio varies, as $SiO_4$ tetrahedra are not affected by the substitution.

Around 1.89 Å, a second feature becomes more apparent with increasing Fe-content. This feature is assigned to Fe-O contributions. This distance is intermediate between those expected for $^{[4]}Fe^{3+}$-O and $^{[5]}Fe^{3+}$-O, and is too small to be assigned to $^{[4]}Fe^{2+}$-O and $^{[5]}Fe^{2+}$–O distances.[10, 24] For NFS sample, this peak consists of two Gaussian components corresponding to Fe-O distances at $1.87 \pm 0.01 Å$ and $2.01 \pm 0.01 Å$. The shorter Fe-O distance is typical for $^{[4]}Fe^{3+}$ and the second has been assigned to $^{[5]}Fe^{3+}$ and $^{[5]}Fe^{2+}$.[10]

The Al-O contribution is expected around 1.75 Å.[30] A Gaussian fit of the first peak in NAS gives $d_{Al\text{-}O}$ = 1.76 ± 0.01 Å and $CN_{Al\text{-}O}$ = 3.9 ± 0.1, in agreement with the presence of $AlO_4$ tetrahedra.[30] Since the neutron weight of this pair is low compared to Si-O or Fe-O (Table 2) and Al-O distance is comprised between Si-O and Fe-O distances, this contribution is not resolved in the glasses of intermediate composition. A shoulder on the low r side of the peak at 2.66 Å, around 2.30 Å, is assigned to Na-O contributions, as in other glasses,[24, 25] including Fe- and Al-bearing silicate glasses.[26, 31]

The third maximum, at 2.66 Å, is characteristic of O-O distances in $SiO_4$ tetrahedra.[27, 32] The intensity of the shoulder located on the high r side of this feature increases with the Al content of the glass. This shoulder is assigned to the contribution of O-O correlations in $AlO_4$ tetrahedra. Indeed, assuming $d_{Al\text{-}O}$ = 1.76 Å and a regular geometry of $AlO_4$ tetrahedra, the O-O contribution of the $AlO_4$ tetrahedra is expected near 2.9 Å.[30] There is no evidence of a further O-O contribution arising from $FeO_4$ tetrahedra and expected at 3.1 Å for a regular site geometry. The absence of this contribution may arise from a distortion of the Fe-sites.

Further features, between 3 and 6 Å, arise from MRO contributions and cannot be unambiguously assigned to atomic pairs. However, the intensity of two contributions, around 3.2 Å and 4.4 Å, increases as the Fe/Al ratio increases. The contribution at 3.2 Å is assigned to a Fe-X pair (X = Si, Fe/Al), the contribution of Fe-Na pair being unlikely due to its low weighting factor and the expected large dispersion of the corresponding distances. The T-second nearest oxygen (T-O(2)) pair correlations (T=Si, Al or Fe), are visible around 4.2-4.4 Å: the feature at 4.2 Å, with a constant intensity is usually assigned to Si-O(2), and at 4.4 Å a feature appears as Fe content increases and can be assigned to Fe-O(2) by comparison with Si-O(2). Finally, the feature around 5.1 Å is assigned to O-O(2) pairs within cationic polyhedra (i.e. $SiO_4$, $AlO_4$ and/or $FeO_x$, x = 4 or 5).

### C. Numerical modeling of short-range order

Additional structural information was obtained by performing EPSR modeling. The experimental and calculated structure factors are presented on Fig. 1. A good agreement is obtained between the experimental and the calculated functions for all glasses along the join. The EPSR-derived partial pair distribution functions (PPDF's) for X-O pairs (X = Fe, Al, Si and Na) are presented in Fig. 4. The average coordination numbers and the contributions of the different coordination numbers to this average (Table 5) have been calculated using cut off distances corresponding to the first minimum in the X-O PPDFs (2.35 Å, 2.67 Å, 2.50 Å and 3.4 Å, ± 0.02 Å, for Si-O, Fe-O, Al-O and Na-O respectively). The X-O PPDF's can be superimposed for all the samples along the join. Neither the interatomic distances, the inter-polyhedral bond angle distributions, nor the coordination numbers of the cations are affected by the substitution of Fe for Al. The short-range organization of the glassy network is then not modified by this substitution. It confirms that the network in the intermediate glasses is the weighted sum of the structures of the two end-members, as assessed from the structure factors.

Si is 4-coordinated in all compositions with 1% or less of 3- and 5-coordinated Si. This indicates a small distribution of Si-O distances and then a small distortion of $SiO_4$ tetrahedra, in agreement with the small value of the Debye-Waller factor obtained by Gaussian fitting of $D_{NFS}(r)$.[10] The average coordination number of Al is also 4, with 5-6 or 2-7% amount of 3- and 5-coordinated Al, respectively. The Al-O distances are then more distributed than Si-O ones, as reflected by the broader first contribution on the Al-O PPDF (full width at half maximum, FWHM = 0.20 Å), as compared to Si-O PPDF (FWHM = 0.16 Å). The $AlO_4$ tetrahedra are then more distorted than the $SiO_4$ ones. The average value of O-Si-O and O-Al-O inter-tetrahedral bond angle distributions, centered on 109°, is in good agreement with the ideal value of 109.4° in regular tetrahedra. The broader distribution of O-Al-O angles compared to O-Si-O angles is also assigned to a larger distortion of the $AlO_4$ tetrahedra. Two Fe-sites are present in the investigated glasses, whatever their Fe content, corresponding to majority $^{[4]}Fe$ with minor contributions of $^{[5]}Fe$. The small amount of $^{[6]}Fe$ (1-4%) arises from computing uncertainties and is considered as $^{[5]}Fe$. According to the decreasing Fe-average coordination number, the relative proportion of $^{[4]}Fe$ compared to $^{[5]}Fe$ increases from NFS to NFA0.5. The O-Fe-O inter-polyhedral bond angle distribution centered on 100° is broad. This indicates the presence of the two Fe-populations and the distortion of $FeO_x$ polyhedra. Moreover, the first peak in $g_{Fe\text{-}O}(r)$ is more asymmetric on high r-side and is broader than in $g_{Al\text{-}O}(r)$ and $g_{Si\text{-}O}(r)$ (FWHM = 0.30 Å for $g_{Fe\text{-}O}(r)$), showing the presence of the two different Fe-O distances revealed by Gaussian fitting of $D_{NFS}(r)$.[10] However, the $^{[5]}Fe$ sites computed by EPSR modeling correspond to either $Fe^{3+}$ or $Fe^{2+}$ and this absence of sensitivity to the valence state can explain the wide distribution of site geometry and $d_{^{[5]}Fe-O}$ distances.

The Na coordination number increases from $CN_{Na\text{-}O}$ = 5.7 to 7.0 with Fe-content, while the Na-O distance remains equal to 2.30 Å in agreement with previous work.[25, 26, 33, 34] This increasing coordination number can reflect a modification in the structural behavior of Na. NAS glass is considered as a fully polymerized three-dimensional glassy network,[35] where Na acts as a charge compensator. The presence of $FeO_5$ polyhedra in the other glasses does not require charge compensation, and then some Na will act as a modifier cation.

In oxide glasses, oxygen atoms play a major role in defining the topology of a network built from rigid tetrahedra. On O-O PPDF's (Fig. 5), the first peak corresponds to the O-O distances within the network-forming tetrahedra. The maximum at 2.60 Å is assigned to the contribution from $SiO_4$ tetrahedra. The influence of the $AlO_4$ tetrahedra is responsible for the slight broadening of this first peak towards larger distances, but EPSR modeling underestimates this contribution of $AlO_4$ tetrahedra (O-O distances expected around 2.9 Å). With increasing Fe content, a contribution appears around 3.1 Å that corresponds to the contribution of O-O linkages in $FeO_x$ (x = 4 or 5) polyhedra. The coordination number of O, i.e. the number of tetrahedra (Si, Al and $^{[4]}$Fe) bound to an oxygen, confirms the presence of majority bridging oxygens, BO ($CN_{O-T}$ = 2), characteristic of the structure of these framework glasses (Fig. 6). The relative proportion of non-bridging oxygens, NBO ($CN_{O-T}$ = 1), increases as Fe increases, due to the increasing amount of $^{[5]}$Fe as the Fe content increases. The proportion of a minority of triclusters, i.e. oxygens linked to three tetrahedra, decreases (from 7 to 3%) as Fe content increases.

### D. Numerical modeling of medium-range order

The calculated cation-cation radial distribution functions (Fig. 7) are identical for all glasses. The Si-Si PPDF's present a first intense and narrow maximum at 3.15 Å, a distance characteristic of corner-sharing $SiO_4$ tetrahedra. The first maximum in the Al-Si and Fe-Si PPDF's appears at 3.20 Å and 3.35 Å, respectively and correspond to $SiO_4$ tetrahedra sharing corners with $AlO_4$ and $FeO_x$ polyhedra, respectively. Such linkages are confirmed by the observation of the simulated structures (Fig. 8). Figure 8 illustrates the homogeneity of the structure of the NAS glass, with $AlO_4$ tetrahedra randomly distributed in the network and sharing corners with $AlO_4$ and $SiO_4$ tetrahedra. As long as Fe is substituted for Al, and even at low Fe-content, $^{[5]}$Fe starts to cluster, as shown on the figure 8 for NFA0.5 glass. In those clusters, $FeO_5$ polyhedra tend to share edges with the other $FeO_5$ polyhedra. By contrast, $^{[4]}$Fe is randomly distributed and shares corners with $SiO_4$ and $AlO_4$ tetrahedra.

The first peak of Fe-Fe PPDF comprises two distinct contributions, at 2.9 Å and 3.4 Å corresponding to the first maximum of $g_{^{[5]}Fe-^{[5]}Fe}(r)$ and $g_{^{[4]}Fe-^{[4]}Fe}(r)$ respectively. The short $^{[5]}$Fe-$^{[5]}$Fe distance corresponds to edge-sharing $FeO_5$, whereas the longer $^{[4]}$Fe-$^{[4]}$Fe distance corresponds to corner-sharing $FeO_4$ tetrahedra. The Fe-Al PPDF is also asymmetric with a first maximum of $g_{Al-^{[5]}Fe}(r)$ at 2.8 Å. As for Fe-Fe PPDF, this contribution at short distances corresponds to $AlO_4$ tetrahedra sharing edges with $FeO_5$ polyhedra. The second maximum at 3.4 Å corresponds to corner-sharing $AlO_4$ and $FeO_4$ tetrahedra. On the contrary, the Al-Al PPDF is narrower and presents only one maximum at 3.25 Å corresponding to corner-sharing $AlO_4$ tetrahedra. These different linkages between $SiO_4$, $AlO_4$, $FeO_4$ and $FeO_5$ polyhedra are interpreted below in terms of different structural behavior of those cations.

The PPDF's involving Na are broad, due to a disordered environment. The first Na-Si distance at 3.2 Å is characteristic of $NaO_x$ polyhedra linked to $SiO_4$ tetrahedra. The first maximums in Na-Fe and Na-Al PPDF's are observed at 3.35 Å and 3.25 Å respectively. They are less broad than in $g_{Na-Si}(r)$, reflecting a less distributed arrangement of Na around Fe and Al, as compared to Si. This may indicate a charge-compensating role of Na near $FeO_4$ and $AlO_4$ tetrahedra.

# IV. DISCUSSION

## A. Differences between Fe- and Al-sites

Our study points out the difference existing between the environment of $Al^{3+}$ and $Fe^{3+}$ in nominally 3D-framework glasses: Al is only 4-coordinated, as $Fe^{3+}$ is 4- and 5-coordinated. The two cations do not have the same role in the silicate network. Only $^{[4]}Al^{3+}$ exists in the NAS glass, this coordination number is lower than in the corresponding crystalline phase, in which Al is only 6-coordinated.[36] This difference between coordination number of Al in $NaAlSi_2O_6$ glass and crystal might contribute to the relative stability of the glass towards devitrification. The $d_{Al-O}$ distance in the glasses investigated is in agreement with the one determined in $AlO_4$ tetrahedra in other silicate glasses.[31, 37]

Only a part of $Fe^{3+}$ is 4-coordinated (~60% of total Fe), the remaining $Fe^{3+}$ and all $Fe^{2+}$ being 5-coordinated. The $FeO_4$ tetrahedra are more distorted than the $AlO_4$ tetrahedra, as shown by the broader feature observed for $^{[4]}$Fe-O contribution in $D_{NFS}(r)$ ($\sigma_{Fe-O}$ = 0.07 Å for the first Gaussian component assigned to $^{[4]}$Fe-O in $D_{NFS}(r)$) as compared to the Al-O contribution in $D_{NAS}(r)$ ($\sigma_{Al-O}$ = 0.05Å for the Al-O component of the Gaussian fit of the first peak of $D_{NAS}(r)$). The first contribution on the $g_{Fe-O}(r)$ can then be assigned to the presence of the two populations of Fe species, which largely overlap due to the distribution of Fe-O distances in the $FeO_x$ (x = 4 or 5) polyhedra. The first Fe population (4-coordinated Fe) corresponds to an average Fe-O distance of 1.87 Å, characteristic of $^{[4]}Fe^{3+}$. The second population (5-coordinated Fe) corresponds to an average distance $d_{Fe-O}$ = 2.01 Å and includes the presence of both $^{[5]}Fe^{2+}$ and $^{[5]}Fe^{3+}$. These higher-coordinated species would act as network modifiers. That can explain that Fe-rich melts have a lower viscosity than their Al counterparts,[16, 17] assuming that glass structure is retained in the molten state in strong liquids, such as silicates.[38] The existence of an important proportion of

$^{[5]}Fe^{2+}$ and $^{[5]}Fe^{3+}$ is in agreement with the structure of the NFS glass.[10,39] In 3D-framework oxide glasses, the surrounding of Fe is similar to that in other silicate glasses, for instance, recent Molecular Dynamics simulations[40] indicate that $^{[4]}Fe^{3+}$ and $^{[5]}Fe^{2+}$ are the most abundant Fe-species, with some additional contribution of $^{[5]}Fe^{3+}$. It turns out that 5-coordinated cations are widespread in oxide glasses, despite the fact that this surrounding is unusual in the crystalline state. Such a coordination number has been shown for several important glass components, including Mg,[41, 42] Al in calcium aluminosilicate glasses,[43] and transition elements (Ti, Fe, Ni) in silicate glasses.[44-48] Depending on the species, the geometry of 5-coordinated cations can range from square-based pyramid, e.g. for $^{[5]}Ti^{4+}$,[45, 46] to trigonal bipyramid, e.g. for $^{[5]}Ni^{2+}$.[47-49] Moreover, different geometries have been suggested for $Fe^{3+}$ and $Ti^{4+}$ from different partial molar volume behavior of $Fe_2O_3$ and $TiO_2$ as a function of composition or of temperature.[50] Our EPSR modeling shows that $FeO_5$ sites correspond to a broad range of distorted polyhedra ranging from trigonal bipyramid to square-based pyramid.

The number of NBO increases as Fe content increases; the network is then less polymerized. This is in agreement with a network former behavior of Al and a mixed role of Fe with a majority of network former $^{[4]}Fe^{3+}$, the remaining Fe, both $^{[5]}Fe^{3+}$ and $^{[5]}Fe^{2+}$, acting as network modifier. Al-O and $^{[4]}$Fe-O bonds being stronger than $^{[5]}$Fe-O ones,[51] Al and $^{[4]}$Fe would reinforce the network.

## B. Cation distribution

The distribution of a cation, Fe (or Al) for example, can be evaluated by determining the ratio between Fe (or Al) next-nearest neighbor (NNN) and the total number of its NNN's (Table 6). In the case of a random distribution, this ratio depends only on the glass stoichiometry. For Fe, in NFS, NFA0.8 and NFA0.5 glasses, EPSR modeling gives ratios that are higher than in the case of a random distribution (Table 6), that implies a trend of Fe to segregate, whatever the Fe-content is. For Al, in NFA0.8, NFA0.5 and NAS glasses, the calculated ratios are close to the theoretical ones, showing that Al is randomly distributed in the silicate framework (Fig. 8), in agreement with the results obtained on charge-balanced aluminosilicates.[52]

We have shown the trend of Fe to an heterogeneous distribution in the glass structure. Besides, the behavior of $^{[4]}$Fe and $^{[5]}$Fe can be asserted by calculating the contributions of $^{[4]}$Fe- and $^{[5]}$Fe-NNN to the total number of NNN around a given Fe atom. It turns out that among the various Fe-species, $^{[4]}$Fe is randomly distributed in the network while only $^{[5]}$Fe tends to an heterogeneous distribution (Table 6), as shown for NFS.[39] Such a trend towards clustering in silicate glasses has been asserted from electron paramagnetic resonance as well as from Mössbauer spectroscopy.[53-55] Moreover, randomly distributed $FeO_4$ share corners with other cationic tetrahedra, $SiO_4$, $AlO_4$ and $FeO_4$ (Fig. 8). This corresponds to the maximum at 3.35 Å in Fe-Si PPDF, and to the second maximum at 3.3 Å and 3.4 Å in Fe-Al and Fe-Fe PPDF's respectively. As Al, $^{[4]}$Fe acts as a network former. EPSR modeling suggests that $FeO_5$ polyhedra tend to share edges (Fig. 8). Such linkages are at the origin of the contribution at 2.9$\AA$ in Fe-Fe PPDF. The clusters imply from 2-3 $FeO_5$ polyhedra in NFA0.5 and up to 5 polyhedra in NFS glass, and they are always linked to the rest of the network either by sharing corners or edges with the cationic tetrahedra ($SiO_4$, $FeO_4$, $AlO_4$). It is then important to note that they do not represent a separated phase. This trend towards $^{[5]}$Fe clustering confirms the presence of domains enriched in network modifier cations, as predicted by the Modified Random Network (MRN).[56]

## C. Role of sodium

The difference in the behavior of Fe and Al passes on the structural role of Na. In silicate glasses, depending on glass composition, Na can act as a network modifier as well as a charge compensator. In the case of aluminosilicate glasses, Na is expected to act as a charge compensator if the ratio Na/Al is smaller or equal to 1; however, if this ratio is greater than 1, the excess of Na shall act as a network modifier.[57] The increase in Na coordination number from 5.7 to 7.0, when Fe content increases, may be interpreted as an indication of a different structural role of Na in Al-bearing and Fe-bearing silicate glasses. In the aluminosilicate end-member (NAS), $Na^+$ acts mainly as a charge compensator to stabilize the negatively charged $AlO_4$ tetrahedra. As Al is substituted with Fe, the proportion of $FeO_5$ increases and Na becomes available as a network modifier. At the same time, the average coordination number of O with tetrahedrally coordinated ions decreases with increasing Fe content: Na atoms, available as network modifier, together with $^{[5]}$Fe are responsible for the depolymerization of the network and then for the formation of NBO.

## D. Influence of Fe-Al substitution on physical properties of sodosilicate glasses

Fe and Al have a different influence on many properties of silicate glasses and melts. For instance, the presence of $Fe_2O_3$, and more particularly $Al_2O_3$, globally increases the chemical durability of sodosilicate glasses.[58] Although different parameters such as pH and chemical composition of the leaching solution, glass texture, or temperature can affect the leaching process,[59] composition has a key role, for example, a charge-compensating role of alkali and alkaline earth cations preventing their diffusion and improving the chemical stability of the glass.[60] $^{[4]}Fe^{3+}$ would then play the same stabilizing role as Al. Contrary to $^{[4]}Al^{3+}$ and $^{[4]}Fe^{3+}$, $^{[5]}Fe$ do not need charge compensation, and acts as a network modifier. In Fe-bearing sodosilicates or aluminosilicates, both $^{[5]}Fe$ and $Na^{+}$ will then be able to diffuse more easily.[15, 61, 62] These two combined effects decrease the chemical durability of Fe-bearing sodosilicates compared to Al-bearing ones.[58]

Tangeman and Lange[63] showed that the configurational heat capacities ($C_P$) of sodium silicate and aluminosilicate liquids are temperature independent, while $C_P$ of Fe-bearing silicate melts shows a negative temperature dependence. This dependence for Fe-bearing silicate melts, has been attributed to the formation of sub-microscopic domains (relatively polymerized and depolymerized) in the Fe-bearing melts that breaks down to a more homogeneous structure with increasing temperature. On the contrary, the aluminosilicate network would be homogeneous whatever the temperature. These observations are in agreement with a homogeneous repartition of $AlO_4$ tetrahedra and trend towards clustering of $FeO_5$ species in the sodosilicate glasses studied in our work.

Viscosity can be considered as an image of the bond strength in the liquid: at a given temperature the stronger the bonds, the more viscous the liquid.[64] During the substitution of Si for Fe (or Al) in charge-compensated compositions, the decrease of viscosity[16] might be due to two phenomena. First, $^{[4]}$Fe-O bonds are weaker than Si-O bonds,[51] the decrease of viscosity can then be explained by the change in bond strengths even if $^{[4]}Fe^{3+}$ acts as a network former. Second, the presence of higher-coordinated (5 and/or 6) Fe, acting as a network modifier, induces the formation of non-bridging oxygens, causing the depolymerization of the network, that further leads to decreasing viscosity, if the glass structure is retained in the liquid state, which is the case for strong liquids such as silicates.[38]

Owing to the dependence of elastic properties of Fe-bearing glasses on the alkali content, a dependence that is different from the ones of aluminosilicates, Burkhard[65] concludes that the structural behavior of Fe and Al is also different, the second one acting as a network former and giving then better elastic properties to the glass. This is consistent with the decrease in the activation energy of viscous flow[66] when Al is substituted for Fe, that has also been assigned to a different structural behavior of Fe and Al. The results presented here are then in agreement with previous observations and allow us to give an explanation to those phenomenon: the presence of 5-coordinated Fe, even if it is a minority species, seems to have a key influence on the physical properties of those materials.

# V. CONCLUSION

The combination of high-resolution neutron and structural modeling using EPSR allowed us to investigate the effects on the structure of Fe/Al substitution in nominally 3D network glasses. The good agreement between the experimental and simulated structure factors and the reproducibility of the modeling allowed an accurate determination of both short- and medium-range organization of the glass structure, pointing out the differences between the structural properties of Fe and Al and the clustering of $^{[5]}Fe$.

The short-range environment around the cations is not affected by the change of composition; the structural behavior of Fe and Al are then not composition dependent. The structure of intermediate glasses (NFA0.8 and NFA0.5) can be seen as a mixture between the configurations represented by the two end-members (NFS and NAS). In all Al-bearing glasses, Al occupies tetrahedra, larger than the $SiO_4$ ones and regularly copolymerized together. Al then acts as a network former, and in this case Na is a charge-compensator of the negatively charged $AlO_4$ tetrahedra. That further confirms the stabilizing role played by Na in building of a 3D polymerized aluminosilicate network. Contrary to Al, Fe occupies also 5-coordinated sites, in addition to a majority of 4-coordinated sites, with both populations playing a different structural role in the network. The high resolution of neutron data allowed determining the average Fe-O distances of these populations. The combination of these experimental data with EPSR modeling gives original structural information on the linking of theses different species. 4-coordinated $Fe^{3+}$ acts as a network former. It is randomly distributed in the network and is regularly connected with $SiO_4$ and $AlO_4$ tetrahedra. By contrast, $FeO_5$ polyhedra tend to segregate even at low Fe-content and to share edges with the other $FeO_5$ sites, a behavior that has been associated with a network-modifying role, causing depolymerization of the network and then formation of non-bridging oxygens. These domains are linked to the rest of the silicate network; this is then not a phase separation. In the presence of Fe, Na acts both as a charge compensator near $AlO_4$ and $FeO_4$ and a modifier that will weaken the 3D network. This difference in the structural behavior of Fe and Al and its consequences on the role of Na, affects the physical properties of Fe- and Al-bearing silicate glasses and melts.

This study highlights the difference between the structural behavior of $Fe^{3+}$ and $Al^{3+}$, which are often considered as having the same role, i.e. network formers in tetrahedral site when their charge is compensated alkalis. It shows that a minority of 5-coordinated Fe can have a large influence on physical properties of glasses.

## ACKNOWLEDGEMENTS

This is IPGP contribution #xxx. The authors acknowledge Catherine McCammon (Bayerisches Geoinstitut, Germany) and Stéphanie Rossano (Laboratoire Géomatériaux et Géologie de l'Ingénieur, France) for Mössbauer spectroscopy.

# TABLES

TABLE I. Experimental glass composition in ato % obtained by electron microprobe analysis, atomic number densities (at.Å$^{-3}$) obtained by Archimedes method and redox ratio obtained by Mössbauer spectroscopy.

| Sample | Fe ± 0.1% | Al ± 0.1% | Si ± 0.2% | Na ± 0.1% | O ± 0.2% | d (at/Å$^3$) ± 0.001 | $Fe^{3+}/Fe_{tot}$ (%) ± 2% |
|---|---|---|---|---|---|---|---|
| NFS | 10.3 | - | 20.4 | 9.7 | 59.6 | 0.072 | 88 |
| NFA0.8 | 8.2 | 2 | 19.8 | 10.1 | 59.9 | 0.072 | 87 |
| NFA0.5 | 5.0 | 5.0 | 19.9 | 10.1 | 60.0 | 0.072 | 86 |
| NAS | - | 10.2 | 19.8 | 10.2 | 59.9 | 0.069 | - |

TABLE II. Neutron weighting factors ($w_{\alpha\beta} = (2-\delta_{\alpha\beta})c_\alpha c_\beta b_\alpha b_\beta$) for each atomic pair α-β in the total structure factors of samples NFS, NFA0.8, NFA0.5 and NAS (eq. 1).

| | Fe-Fe | Fe-Al | Fe-Si | Fe-Na | Fe-O | Al-Al | Al-Si |
|---|---|---|---|---|---|---|---|
| NFS | 0.0097 | - | 0.0167 | 0.0068 | 0.0682 | - | - |
| NFA0.8 | 0.0061 | 0.0011 | 0.0128 | 0.0056 | 0.0541 | 0.00005 | 0.0011 |
| NFA0.5 | 0.0023 | 0.0017 | 0.0079 | 0.0035 | 0.0335 | 0.0003 | 0.0029 |
| NAS | - | - | - | - | - | 0.0012 | 0.0058 |

| | Al-Na | Al-O | Si-Si | Si-Na | Si-O | O-O |
|---|---|---|---|---|---|---|
| NFS | - | - | 0.0072 | 0.0059 | 0.0586 | 0.1196 |
| NFA0.8 | 0.0005 | 0.0048 | 0.0068 | 0.0060 | 0.0572 | 0.1209 |
| NFA0.5 | 0.0013 | 0.0121 | 0.0068 | 0.0060 | 0.0575 | 0.1211 |
| NAS | 0.0026 | 0.0244 | 0.0067 | 0.0060 | 0.0571 | 0.1207 |

TABLE III. Composition of cubic boxes used for EPSR simulations, dimension of the boxes.

| | $Fe^{3+}$ | $Al^{3+}$ | $Si^{4+}$ | $Na^{+}$ | $O^{2-}$ | a (Å) |
|---|---|---|---|---|---|---|
| NFS | 400 | 0 | 800 | 400 | 2400 | 37.9821 |
| NFA0.8 | 320 | 80 | 800 | 400 | 2400 | 38.1571 |
| NFA0.5 | 200 | 200 | 800 | 400 | 2400 | 38.1571 |
| NAS | 0 | 400 | 800 | 400 | 2400 | 38.7023 |

TABLE IV. Parameters for the starting potential in the EPSR simulations.

| | Coulomb charges | ε (kJ/mole) | σ (Å) |
|---|---|---|---|
| $Fe^{3+}$ | +1.5 e | 0.15 | 1.7 |
| $Al^{3+}$ | +1.5 e | 0.26 | 1.26 |
| $Si^{4+}$ | +2 e | 0.175 | 1.06 |
| $Na^{+}$ | +0.5 e | 0.175 | 2.1 |
| $O^{2-}$ | -1 e | 0.1625 | 3.6 |

TABLE V. Average coordination number obtained using EPSR, and distribution of each coordination number for each species.

| sample | average coord. | % 3-coord | % 4-coord | % 5-coord | % 6-coord | % 7-coord | % 8-coord | % 9-coord | % 10-coord. |
|---|---|---|---|---|---|---|---|---|---|
| **NFS** | | | | | | | | | |
| $CN_{Fe\text{-}O}$ | 4.43 | 1 | 59 | 36 | 4 | 0 | 0 | 0 | 0 |
| $CN_{Si\text{-}O}$ | 4.00 | 0 | 100 | 0 | 0 | 0 | 0 | 0 | 0 |
| $CN_{Na\text{-}O}$ | 7.03 | 0 | 2 | 9 | 23 | 29 | 24 | 10 | 3 |
| std. Dev $CN_{Fe\text{-}O}$ | 0.03 | 0.34 | 2.29 | 2.25 | 0.91 | 0.00 | 0.00 | 0.00 | 0.00 |
| **NFA0.8** | | | | | | | | | |
| $CN_{Fe\text{-}O}$ | 4.35 | 1 | 64 | 33 | 2 | 0 | 0 | 0 | 0 |
| $CN_{Al\text{-}O}$ | 3.97 | 6 | 92 | 2 | 0 | 0 | 0 | 0 | 0 |
| $CN_{Si\text{-}O}$ | 4.00 | 1 | 99 | 0 | 0 | 0 | 0 | 0 | 0 |
| $CN_{Na\text{-}O}$ | 6.77 | 2 | 5 | 18 | 26 | 28 | 14 | 5 | 1 |
| std. Dev $CN_{Fe\text{-}O}$ | 0.02 | 0.00 | 0.79 | 1.56 | 1.44 | 0.00 | 0.00 | 0.00 | 0.00 |
| **NFA0.5** | | | | | | | | | |
| $CN_{Fe\text{-}O}$ | 4.34 | 1 | 66 | 32 | 1 | 0 | 0 | 0 | 0 |
| $CN_{Al\text{-}O}$ | 4.00 | 5 | 90 | 5 | 0 | 0 | 0 | 0 | 0 |
| $CN_{Si\text{-}O}$ | 4.00 | 0 | 100 | 0 | 0 | 0 | 0 | 0 | 0 |
| $CN_{Na\text{-}O}$ | 6.70 | 1 | 3 | 14 | 28 | 29 | 18 | 6 | 2 |
| std. Dev $CN_{Fe\text{-}O}$ | 0.04 | 0.29 | 3.01 | 2.75 | 0.29 | 0.00 | 0.00 | 0.00 | 0.00 |
| **NAS** | | | | | | | | | |
| $CN_{Al\text{-}O}$ | 4.01 | 6 | 87 | 7 | 0 | 0 | 0 | 0 | 0 |
| $CN_{Si\text{-}O}$ | 4.00 | 1 | 99 | 1 | 0 | 0 | 0 | 0 | 0 |
| $CN_{Na\text{-}O}$ | 5.68 | 4 | 13 | 29 | 29 | 16 | 7 | 2 | 0 |

TABLE VI. Fe (or Al, [4]Fe and [5]Fe) next-nearest neighbor (NNN) and the total number of its NNN's calculated from EPSR simulations. The values of those ratios in the case of a random distribution are indicated between brackets.

| | NFS | NFA0.8 | NFA0.5 | NAS |
|---|---|---|---|---|
| $\frac{CN_{Fe-Fe}}{CN_{Fe-Fe}+CN_{Fe-Si}+CN_{Fe-Al}}$ | 0.42 (0.33) | 0.32(0.27) | 0.21(0.17) | - |
| $\frac{CN_{Al-Al}}{CN_{Al-Al}+CN_{Al-Si}+CN_{Al-Fe}}$ | - | 0.06(0.07) | 0.16(0.17) | 0.35(0.33) |
| $\frac{CN_{^{[4]}Fe-^{[4]}Fe}}{CN_{^{[4]}Fe-^{[4]}Fe}+CN_{^{[4]}Fe-^{[5]}Fe}+CN_{^{[4]}Fe-Si}+CN_{^{[4]}Fe-Al}}$ | 0.23(0.20) | 0.19(0.17) | 0.13(0.11) | - |
| $\frac{CN_{^{[5]}Fe-^{[5]}Fe}}{CN_{^{[5]}Fe-^{[5]}Fe}+CN_{^{[5]}Fe-^{[4]}Fe}+CN_{^{[5]}Fe-Si}+CN_{^{[5]}Fe-Al}}$ | 0.23(0.13) | 0.13(0.09) | 0.10(0.05) | - |

## FIGURE CAPTION

FIG. 1. Experimental neutron structure factors (solid lines) and fit to the data (dots) obtained after empirical potential structure refinement. The feedback factor was taken equal to 0.75 for all samples. Curves have been displaced vertically for clarity. The isobestic points are indicated as vertical solid lines in the insert.

FIG. 2. Position of the first peak of F(Q)s as a function of Fe content. The first peak was fitted using a Gaussian based on its low Q side a horizontal background.

FIG. 3. Differential correlation functions obtained by Fourier Transform of the total structure factors. The F(Q) were modified by a Lorch function to reduce the termination effects of the F.T., and Fourier Transformed with a data interval of 0.4-35$Å^{-1}$. Curves have been displaced vertically for clarity.

FIG. 4. (Color online) Cation-oxygen partial pair distribution functions extracted from EPSR simulations for NFA0.8 glass sample (similar functions are obtained for the other glass compositions). Curves have been displaced vertically for clarity.

FIG. 5. (Color online) Oxygen-oxygen partial pair distribution functions extracted from EPSR simulations for the four glasses.

FIG. 6. Oxygen coordination number within tetrahedral species ( Si, Al, $^{[4]}Fe$) calculated form EPSR atomic configurations.

FIG. 7. (Color online) Cation-cation partial correlation functions extracted from EPSR simulations for NFA0.8 glass sample (similar functions are obtained for the other glass compositions). Curves have been displaced vertically for clarity.

FIG. 8. (Color online) Slices containing ~500 atoms (19 Å × 19 Å × 19 Å) into the EPSR configurations of the four glasses. $SiO_4$, $AlO_4$, $FeO_4$ tetrahedra are represented in blue, green and black respectively. $FeO_5$ polyhedra and Na atoms are represented in pink and yellow respectively.

# FIGURE 1

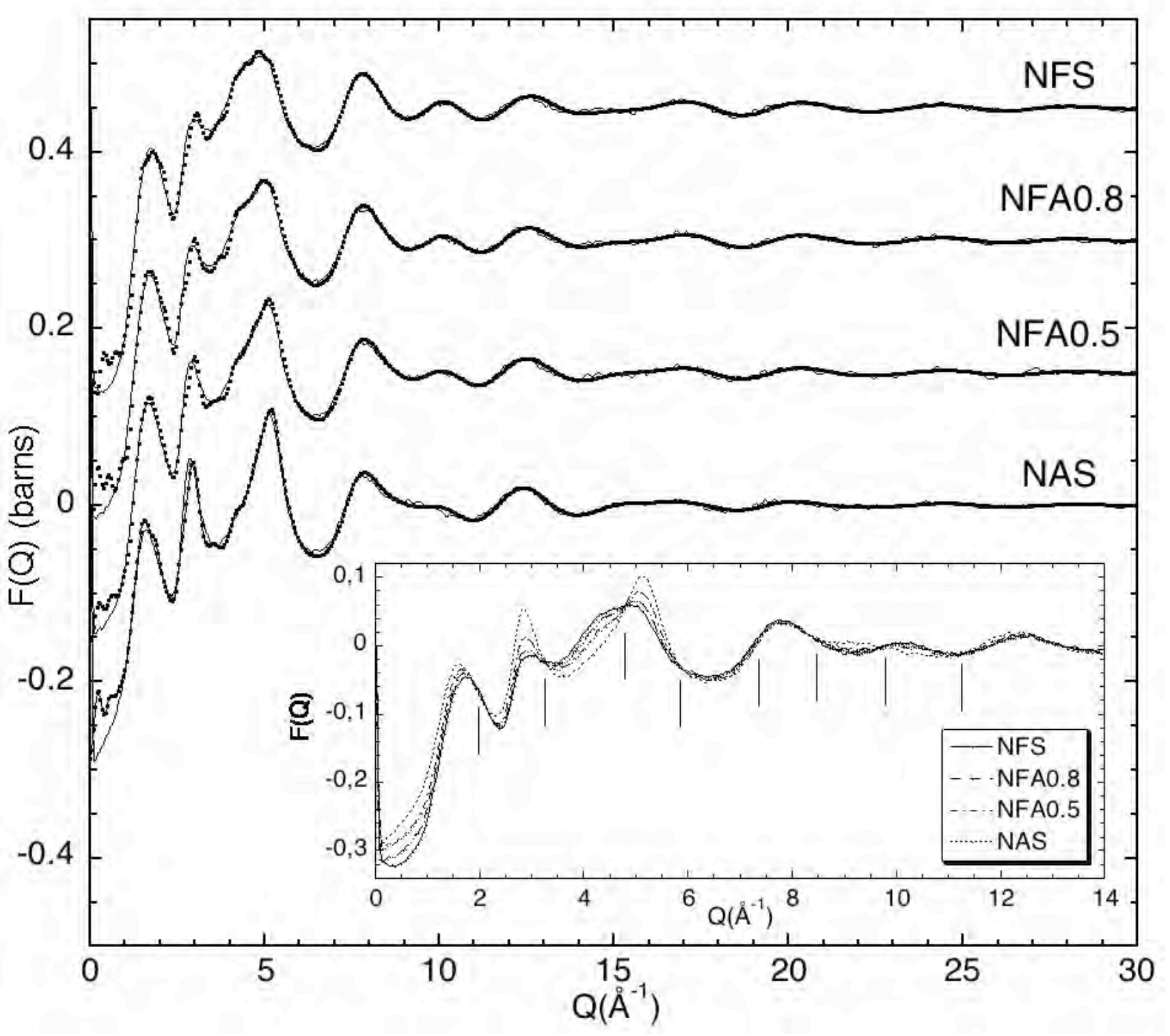

FIG. 1. Experimental neutron structure factors (solid lines) and fit to the data (dots) obtained after empirical potential structure refinement. The feedback factor was taken equal to 0.75 for all samples. Curves have been displaced vertically for clarity. The isobestic points are indicated as vertical solid lines in the insert.

**FIGURE 2**

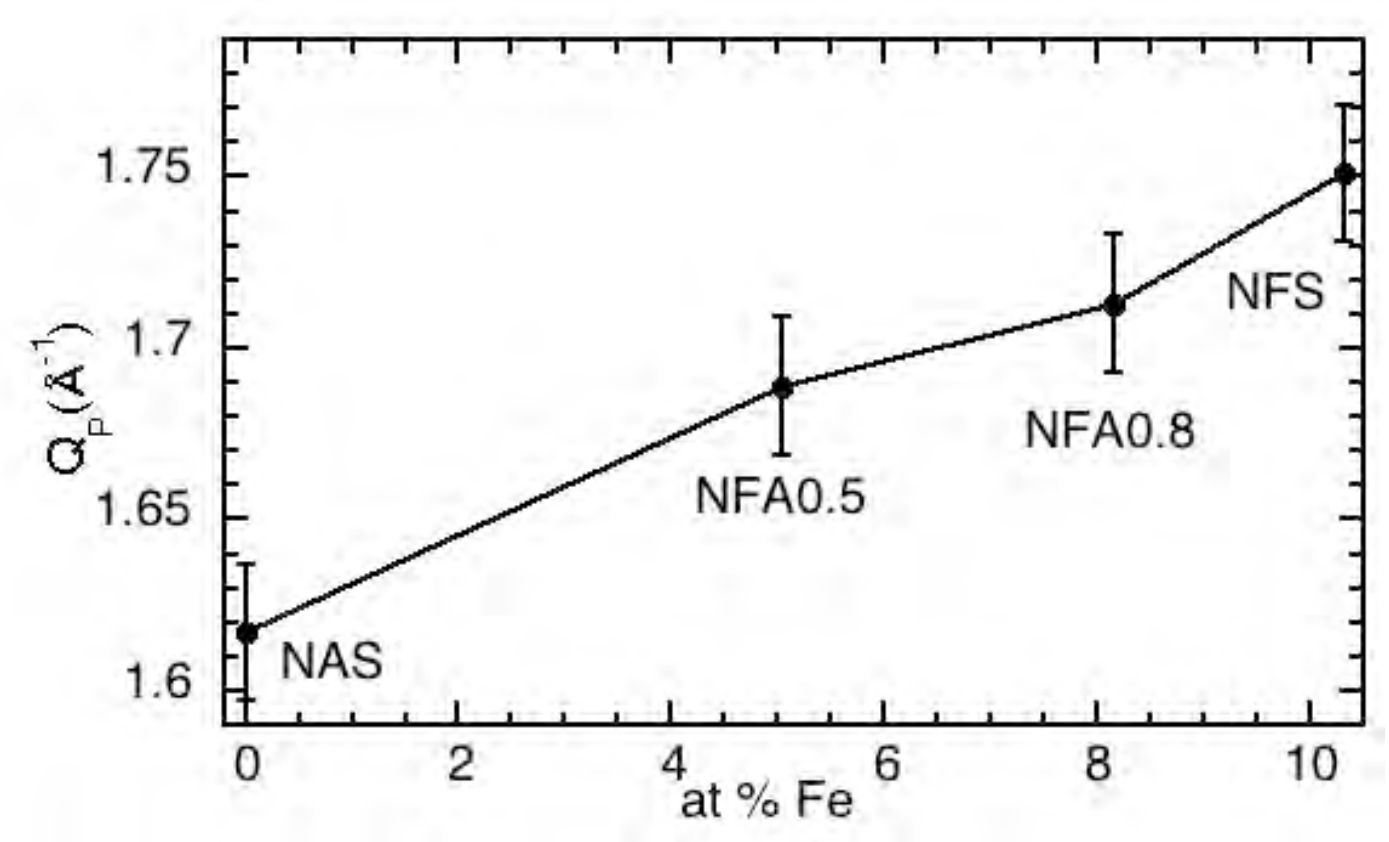

FIG. 2. Position of the first peak of F(Q)s as a function of Fe content. The first peak was fitted using a Gaussian based on its low Q side a horizontal background.

## FIGURE 3

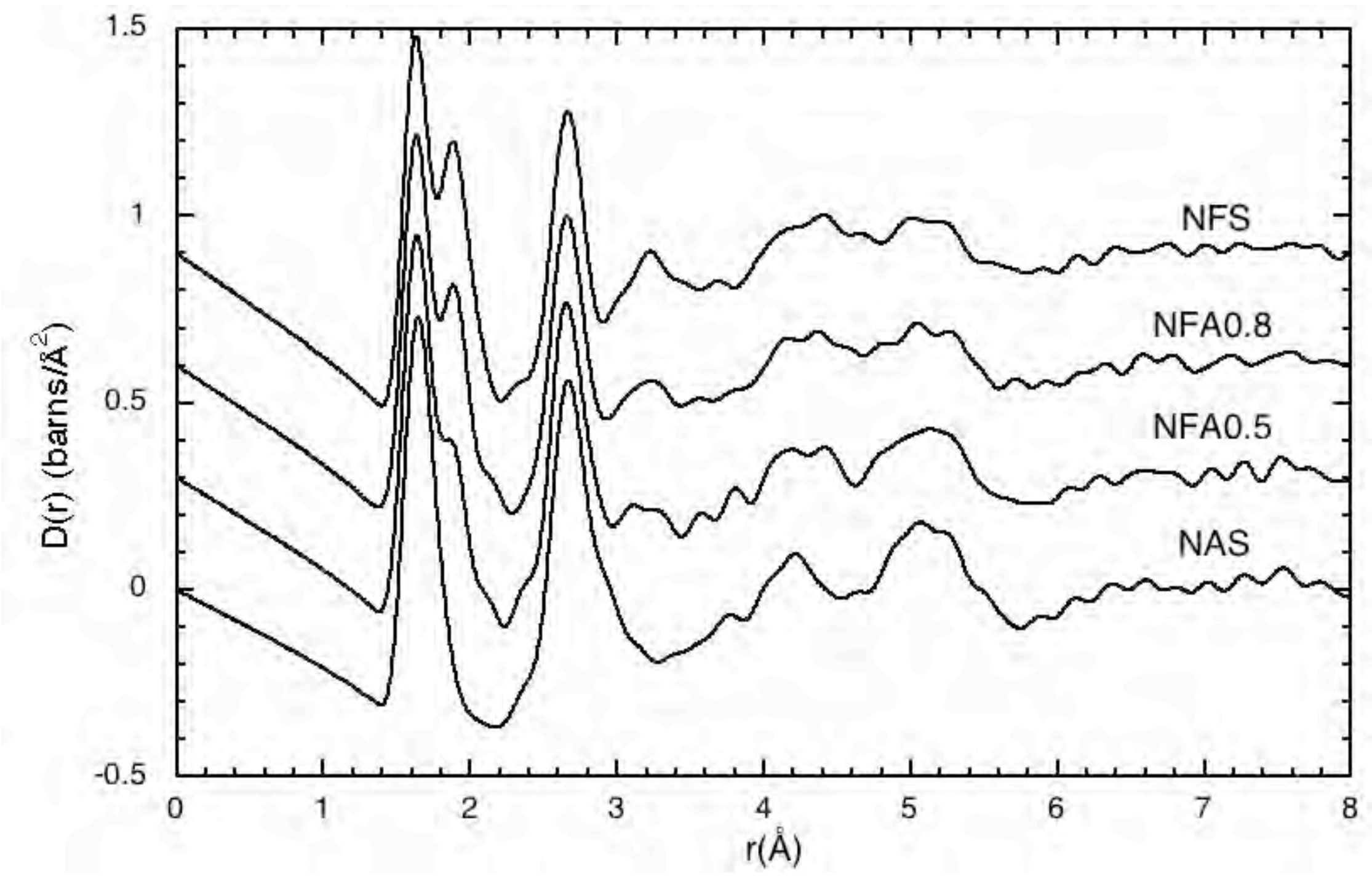

FIG. 3. Differential correlation functions obtained by Fourier Transform of the total structure factors. The F(Q) were modified by a Lorch function to reduce the termination effects of the F.T., and Fourier Transformed with a data interval of 0.4-35Å$^{-1}$. Curves have been displaced vertically for clarity.

**FIGURE 4**

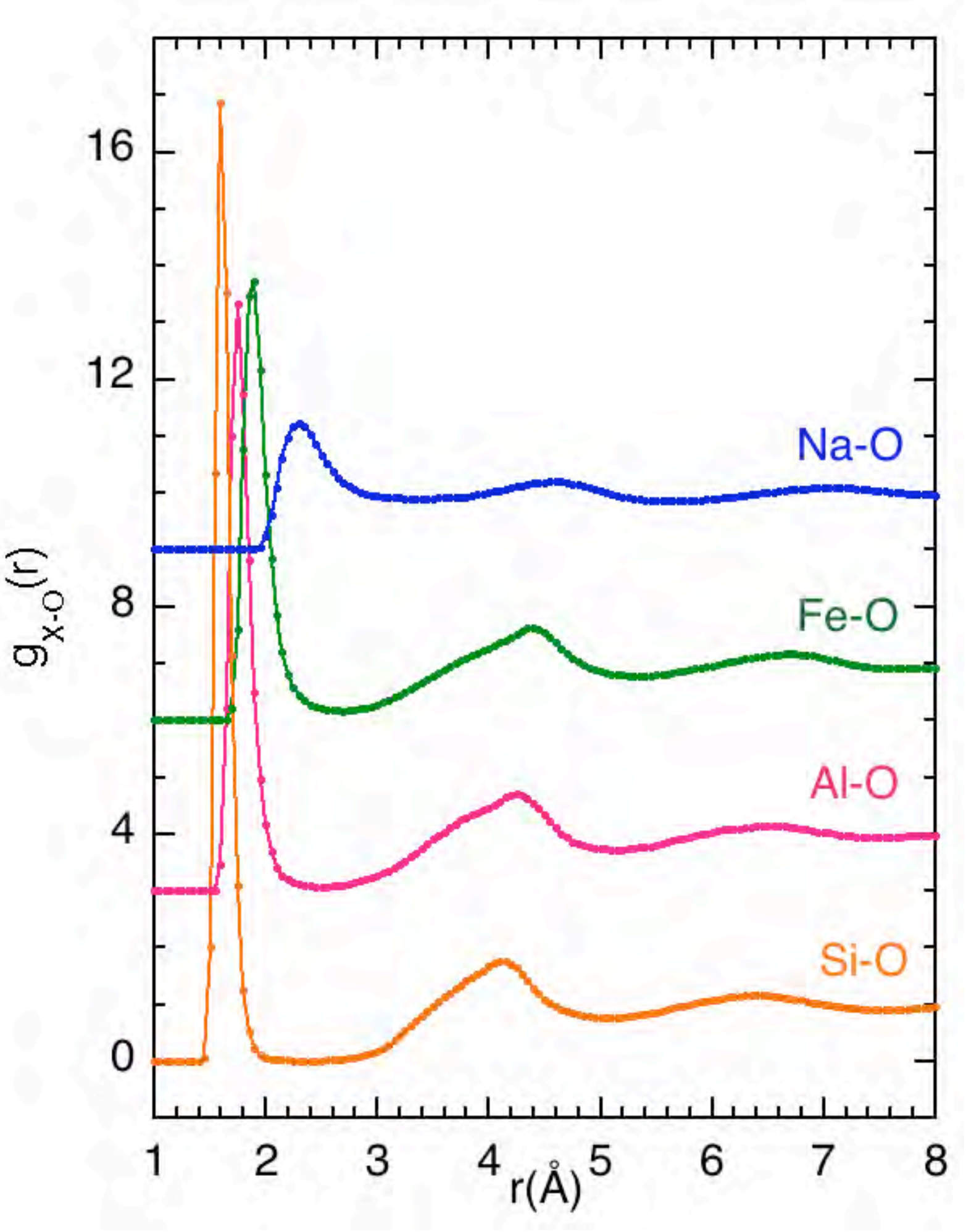

FIG. 4. (Color online) Cation-oxygen partial pair distribution functions extracted from EPSR simulations for NFA0.8 glass sample (similar functions are obtained for the other glass compositions). Curves have been displaced vertically for clarity.

**FIGURE 5**

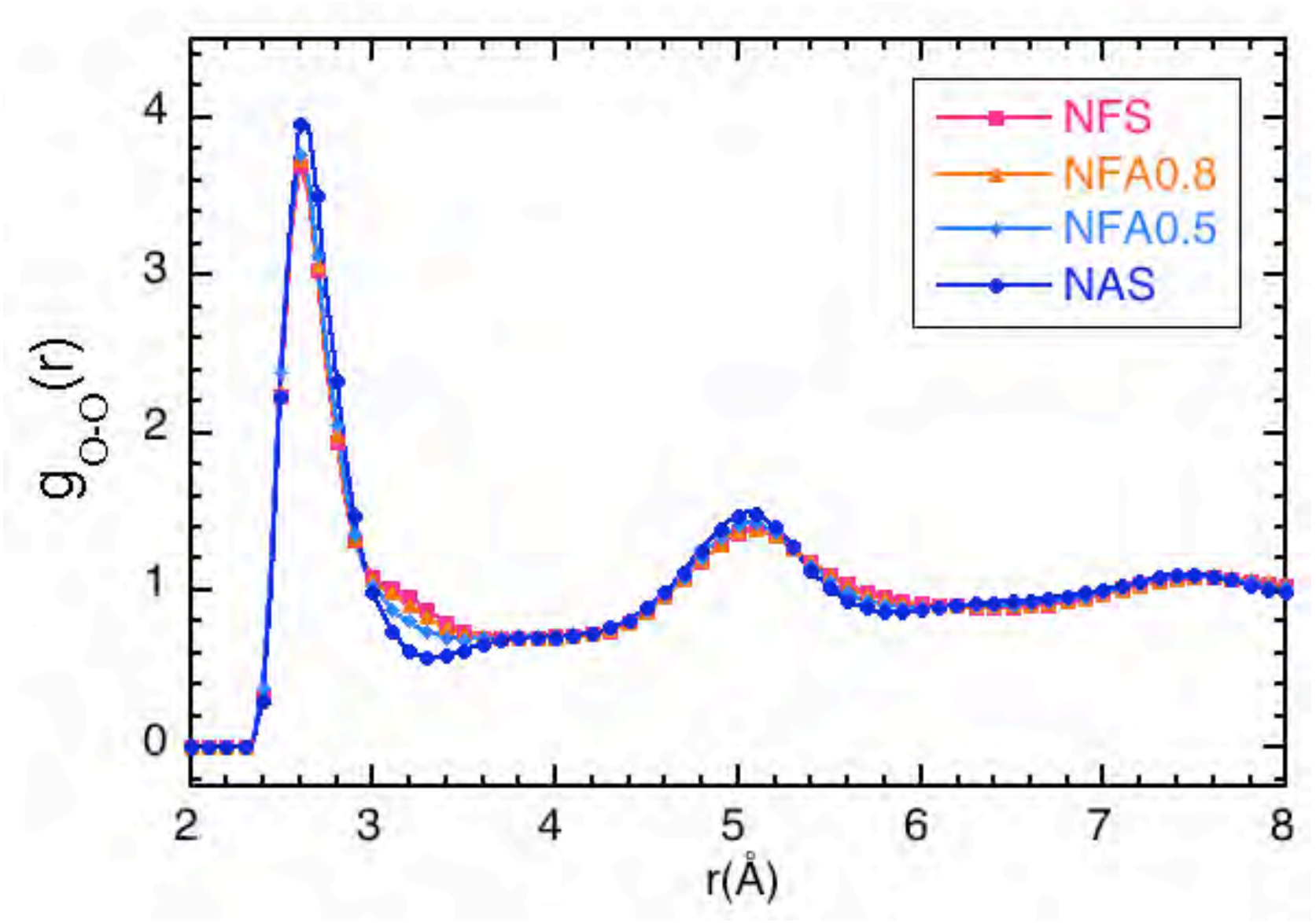

FIG. 5. (Color online) Oxygen-oxygen partial pair distribution functions extracted from EPSR simulations for the four glasses.

**FIGURE 6**

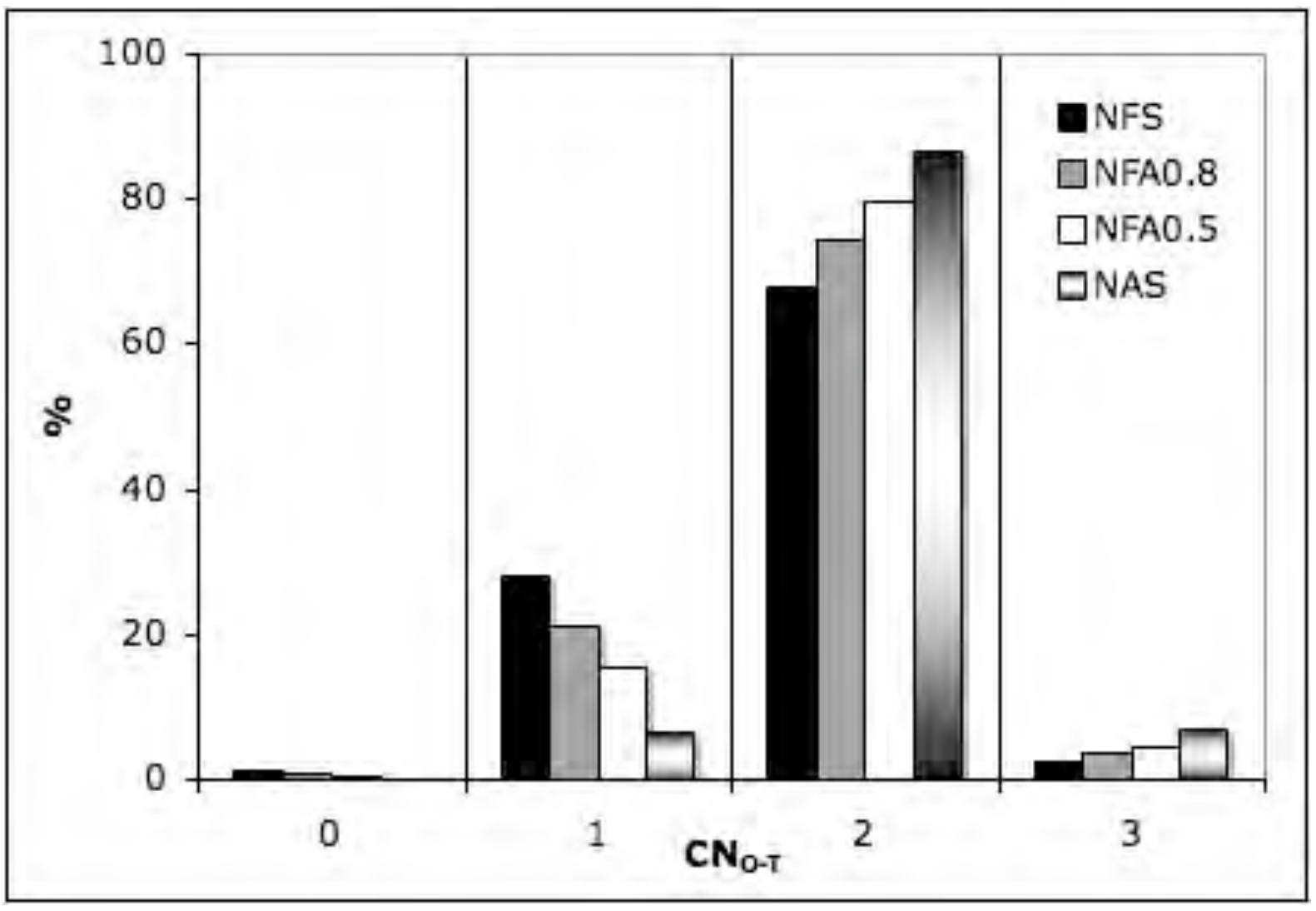

FIG. 6. Oxygen coordination number within tetrahedral species ( Si, Al, [4]Fe) calculated form EPSR atomic configurations.

**FIGURE 7**

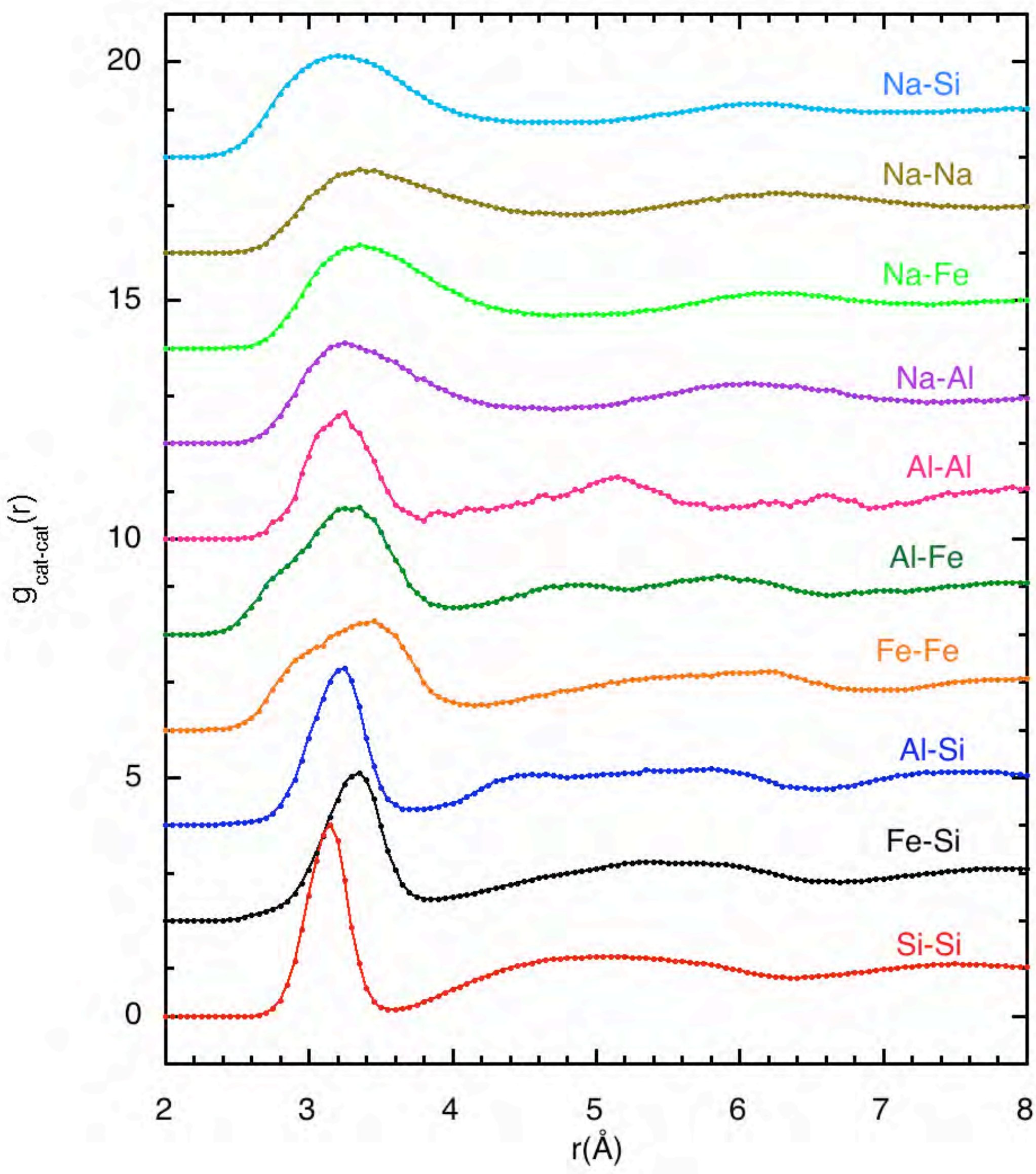

FIG. 7. (Color online) Cation-cation partial correlation functions extracted from EPSR simulations for NFA0.8 glass sample (similar functions are obtained for the other glass compositions). Curves have been displaced vertically for clarity.

**FIGURE 8**

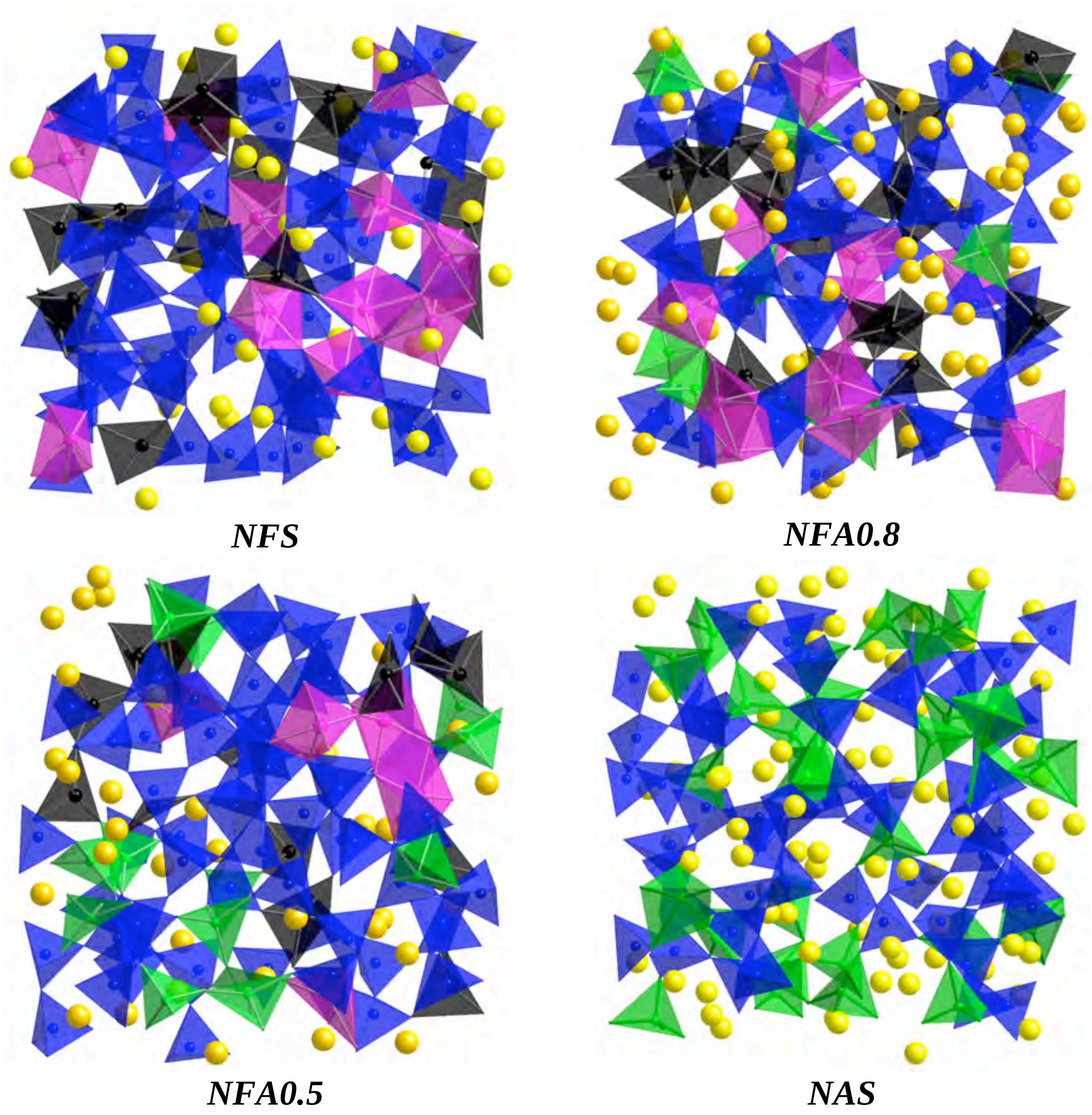

FIG. 8. (Color online) Slices containing ~500 atoms (19 Å × 19 Å × 19 Å) into the EPSR configurations of the four glasses. $SiO_4$, $AlO_4$, $FeO_4$ tetrahedra are represented in blue, green and black respectively. $FeO_5$ polyhedra and Na atoms are represented in pink and yellow respectively.